\documentclass[aps,nofootinbib,reprint,nobibnotes,showpacs,superscriptaddress]{revtex4-2} 

\usepackage[T1]{fontenc}
\usepackage[english]{babel}
 \addto\captionsenglish{
                       }

\usepackage{graphicx}
\usepackage{multirow}

\usepackage[caption=false]{subfig}  
\usepackage[export]{adjustbox}

\usepackage{amsmath}
\usepackage{amssymb}
\usepackage[version=3]{mhchem}
\usepackage{mathrsfs}
\usepackage{bm}

\usepackage{upgreek}

\usepackage[hidelinks,linktocpage,colorlinks]{hyperref}
\hypersetup{
            colorlinks,
            linkcolor={red!70!black}  ,
            citecolor={green!70!black},
            urlcolor ={blue!70!black}
           }

\usepackage{comment} 
\usepackage[switch]{lineno} 
\usepackage[usenames,dvipsnames]{xcolor}
\usepackage{xspace}

\newcommand{\mTa}{\ce{^{180m}Ta}\xspace}
\newcommand{\Ta} {\ce{Ta}\xspace}

\newcommand{\ULBHPGe}{ULB-HPGe\xspace}

\newcommand{\EC}   {EC\xspace}
\newcommand{\IC}   {IC\xspace}
\newcommand{\betaM}{$\beta^-$\xspace}
\newcommand{\tEC}  {\ensuremath{T_{1/2,~\mathrm{EC}}}}
\newcommand{\tb}   {\ensuremath{T_{1/2,~\beta^-}}}
\newcommand{\tECb} {\ensuremath{T_{1/2,~\mathrm{EC}\,+\,\beta^-}}}
\newcommand{\tIC}  {\ensuremath{T_{1/2,~\mathrm{IC}}}}
\newcommand{\tGen} {\ensuremath{T_{1/2}}}

\newcommand{\ckeVkgday}{\ensuremath{\text{c}~\text{keV}^{-1}~\text{kg}^{-1}~\text{d}^{-1}}\xspace}
\newcommand{\mum}   {\ensuremath{\upmu\text{m}}\xspace}
\newcommand{\mummin}{\ensuremath{\upmu\text{m min}^{-1}}\xspace}
\newcommand{\cmsqs} {\ensuremath{\text{cm}^{-2}\,\text{s}^{-1}}\xspace}
\newcommand{\ia}    {\ensuremath{\text{i.\,a.}}\xspace}
\newcommand{\measT} {\ensuremath{t}\xspace}
\newcommand{\gyr }  {\ensuremath{\text{g yr}}\xspace}
\newcommand{\kgyr}  {\ensuremath{\text{kg yr}}\xspace}
\newcommand{\uBqkg} {\ensuremath{\upmu\text{Bq kg}^{-1}}\xspace}

\newcommand{\tableNoteI} {$^{\mbox{\scriptsize $\,\dag$}}$}
\newcommand{\tableNoteII}{$^*$}

\newcommand{\AMcDinst}{\affiliation{Arthur B.\ McDonald Canadian Astroparticle Physics Research Institute, K7L-3N6 Kingston, Ontario, Canada}}
\newcommand{\GSSI}    {\affiliation{Gran Sasso Science Institute, 67100 L'Aquila, Italy}}
\newcommand{\INFNmib} {\affiliation{INFN -- Sezione di Milano--Bicocca, 20126 Milano, Italy}}
\newcommand{\INFNlnf }{\affiliation{INFN -- Laboratori Nazionali di Frascati, 00044 Frascati (Roma), Italy}}
\newcommand{\INFNlngs}{\affiliation{INFN -- Laboratori Nazionali del Gran Sasso, Assergi, 67100 L'Aquila, Italy}}
\newcommand{\INFNroma}{\affiliation{INFN -- Sezione di Roma, 00185 Roma, Italy}}
\newcommand{\LBL}     {\affiliation{Nuclear Science Division, Lawrence Berkeley National Laboratory, Berkeley, CA 94720, USA}}
\newcommand{\QueensU} {\affiliation{Department of Physics, Enigneering Physics and Astronomy, Queen's University, K7L-3N6 Kingston, Ontario, Canada}}
\newcommand{\UniMib}  {\affiliation{Università di Milano--Bicocca, 20126 Milano, Italy}}
\newcommand{\corrAuth}{\email{matthias.laubenstein@lngs.infn.it}}

\begin{document}

 \title{Deep-underground search for the decay of \mTa \\[3pt] with an ultra-low-background HPGe detector}

 \author{R.~Cerroni}    \INFNlngs
 \author{S.~Dell'Oro}   \UniMib   \INFNmib
 \author{A.~Formicola}  \INFNroma
 \author{S.~Ghislandi}  \GSSI     \INFNlngs
 \author{L.~Ioannucci}  \INFNlnf
 \author{M.~Laubenstein}\corrAuth \INFNlngs
 \author{B.~Lehnert}    \LBL
 \author{S.~Nagorny}    \QueensU  \AMcDinst
 \author{S.~Nisi}       \INFNlngs
 \author{L.~Pagnanini}  \GSSI     \INFNlngs

 \date{\today}
   
 \begin{abstract}
  \mTa is the longest-lived metastable state presently known. Its decay has not been observed yet.
  In this work, we report a new result on the decay of \mTa obtained with a $2015.12$-g tantalum sample measured for $527.7$ d with an ultra-low background HPGe detector in the STELLA laboratory
  of the Laboratori Nazionali del Gran Sasso (LNGS), in Italy. Before the measurement, the sample has been stored deep-under\-ground for ten years, resulting in subdominant background contributions from cosmogenically activated \ce{^{182}Ta}.
  We observe no signal in the regions of interest and set half-life limits on the process for the two channels \EC and \betaM:
  $\tEC > 1.6 \times 10^{18}$ yr and $\tb > 1.1\times 10^{18}$~yr ($90$\% C.\,I.), respectively.
  We also set the limit on the $\gamma$ de-excitation~/~\IC channel: $\tIC > 4.1 \times 10^{15}$~yr ($90$\% C.\,I.).
  These are, as of now, the most stringent bounds on the decay of \mTa worldwide. 
  Finally, we test the hypothetical scenarios of de-excitation of \mTa by cosmological Dark Matter and constrain new parameter space for strongly-interacting dark-matter particle with mass up to $10^5$~GeV. 
  \\[+9pt]
  Published on: Eur.\ Phys.\ J.\ C {\bf 83}, 925 (2023) \hfill DOI: \href{https://doi.org/10.1140/epjc/s10052-023-12087-5}{10.1140/epjc/s10052-023-12087-5}
 \end{abstract}

 \maketitle

 \section{Introduction}
 
 Tantalum (\Ta) has a nearly mono-isotopic composition: the natural abundances are in fact $0.9998799(32)$ of \ce{^{181}Ta} and $0.0001201(32)$ of \mTa, while the ground state of \ce{^{180}Ta} is not present, having a half-life of $8.15$~h \cite{Meija:2016iupac}.
 The tiny fraction of \mTa was first measured in 1955~\cite{White:1955aaa,White:1956aaa}, although at that time it was not understood that it belonged to an isomeric state. 
 Since then, significant interest has been demonstrated in probing the half-life of \mTa, given the rather unique situation in nature and the impact in astrophysics. Indeed the half-life of \mTa directly affects its observed abundance, and hence its production mechanism, which in turn is a probe for the production of heavy elements in stellar nucleosynthesis~\cite{deLaeter:2005kc,Mohr:2006wn,Malatji:2019xrs}.
 
\begin{figure}[t]
  \centering
  \includegraphics[width=1.\columnwidth]{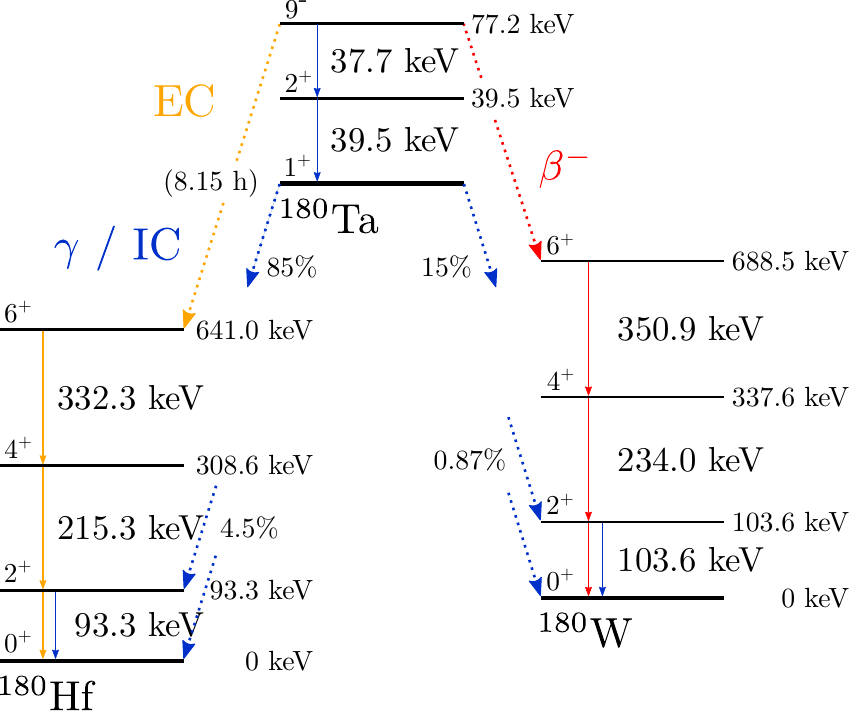}
  \caption{Decay scheme of \mTa~\cite{McCutchan:2015aaa}.
           The \EC, \betaM and $\gamma$ / \IC branches are shown in different colors, while the experimental signatures for $\gamma$ spectroscopy are highlighted.
          }
  \label{fig:decay_scheme}
 \end{figure}

 The decay scheme of \mTa is shown in Fig.~\ref{fig:decay_scheme}. Two possible branches lead to \ce{^{180}W} and to \ce{^{180}Hf} via electron capture (\EC) and \betaM decay, respectively,
 while a $\gamma$ de-excitation / internal conversion (\IC) from the $9^-$ isomeric state to the $2^+$ state is also possible. Recent calculations of the nuclear matrix elements for the transition give an estimate for the half-life of $1.4 \times 10^{20}$ yr (\EC), $5.4 \times 10^{23}$ yr (\betaM) and $8 \times 10^{18}$ yr (\IC), dominated by the latter channel~\cite{Ejiri:2017dro}. It is worth noting that the Authors of Ref.~\cite{Ejiri:2017dro} hint for an overall factor 2 of uncertainty for their calculations and  push for further experimental measurements.

\begin{figure}[t]
  \centering
  \includegraphics[width=1.\columnwidth]{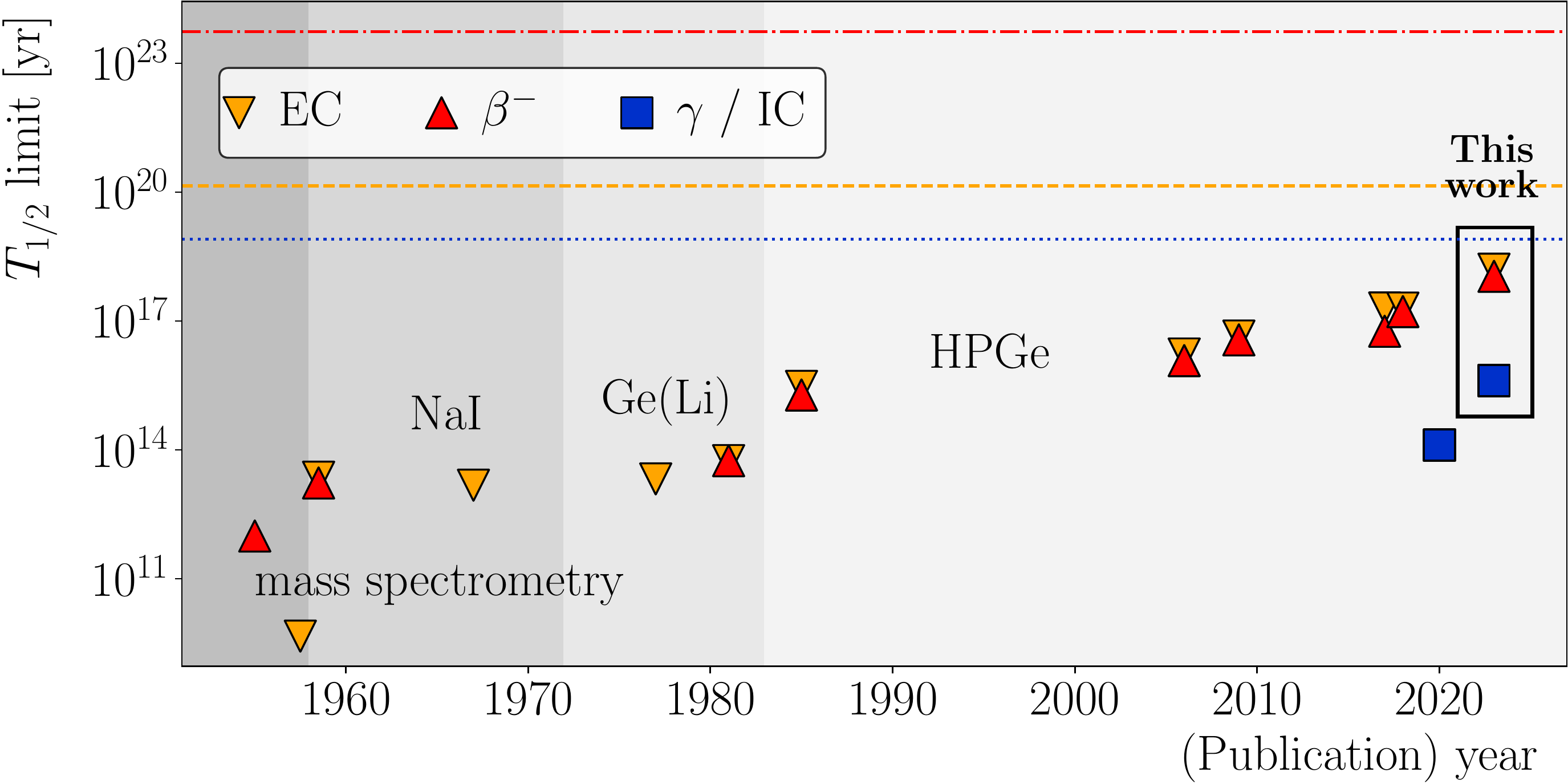}
  \caption{Lower limits on the decay half-life of \mTa on the \EC, \betaM and $\gamma$ / \IC channels~           \cite{Eberhardt:1955aaa,Eberhardt:1958aaa,Bauminger:1958aaa,Sakamoto:1967aaa,Ardisson:1977aaa,Norman:1981aaa,Cumming:1985aaa,Hult:2006aaa,Hult:2009aaa,Lehnert:2016iku,Chan:2018chm,Lehnert:2019tuw}; the box encloses the new results presented in this work.
           The labels and corresponding shadowed areas refer to the different techniques used for the measurements.
           The horizontal lines indicate the theoretical half-lives from the calculations of the nuclear matrix elements~\cite{Ejiri:2017dro}:
           \EC (dashed), \betaM (dash-dotted) and $\gamma$ / \IC (dotted).
          }
  \label{fig:limits}
 \end{figure}

 Over the years, different techniques have been used to experimentally search for the decay of \mTa (Fig.~\ref{fig:limits}).
 The early limits set by mass spectrometry were soon overcome by the more powerful $\gamma$-spectrometry measurements, initially performed with scintillators and later with
 germanium \ce{Ge(Li)} detectors; more recently, stringent results have been obtained by using High-Purity Germanium (HPGe) detectors.
 As of today, neither of the radioactive decays has been observed and the current overall limit on the process is $\tGen > 9.03 \times 10^{16}$ yr at $90\%$ C.\,L.~\cite{chan:2017thesis,Chan:2018chm}.%
 \footnote{This result only considers the combination of EC and $\beta^-$ decay and is reported as preliminary in the cited proceedings.}
 
 In this work, we present new results on the search for the decay of \mTa, where we exploit the excellent performance of the ultra-low-background HPGe (\ULBHPGe) detectors in the
 SubTerranean Low-Level Assay (STELLA) laboratory \cite{Arpesella:1996,Laubenstein:2017yjj} at Laboratori Nazionali del Gran Sasso, in Italy.
 The average overburden of 3600~m w.\,e.\ and the mostly-calcareous rock composition of the Gran Sasso mountain~\cite{Catalano:1986M} guarantee very low muon and neutron fluxes of
 about $3\times 10^{-8}$ \cmsqs~\cite{Ambrosio:1995cx,Borexino:2012wej} and $4\times 10^{-6}$ \cmsqs~\cite{Best:2015yma}, as well as a relative low content of natural radioactivity in the surrounding rock.
 In addition to the deep-underground location, the strict protocols adopted in order to select only radio-clean materials for the detectors and shield parts result
 in a strong abatement of the background due to internal and environmental radioactive content.  
 
 \section{Sample description and preparation}
 
 The sample was procured in 2009 and consists of 6 tiles of metallic \Ta produced via vacuum melting by Advent Research Material Ltd.
 Each tile measures $10\times 10 \times 0.2$ cm$^3$ and has a mass of $\sim 335$ g, corresponding to a total of about $2$ kg.
 
 In order to mitigate the intrinsic background due to cosmogenic activation of \ce{^{182}Ta} ($\tGen = 114.74$ d~\cite{Singh:2015A182}), the tiles have been stored deep-underground in the STELLA laboratory until the start of measurements in 2019. This corresponds to about $30$ half-lives and reduced the \ce{^{182}Ta} background contribution to subdominant levels.
 
 Before starting the measurement campaign, the \Ta sample underwent a chemical treatment aimed at reducing the surface contamination.
 We deemed that removing the outermost $5$ \mum (per side) of each tile would represent a good compromise between an effective cleaning and an affordable mass loss.
 We tested different acid mixtures to assess the corresponding etching action.
 We observed that pure \ce{HF} was not powerful, as it resulted in a $0.02$ \mummin erosion, while adding a small amount of \ce{HNO_3} and \ce{H_2SO_4},
 easily increased the erosion speed to more than $60$ \mummin.
 
 In the end, we adopted the composition: \ce{HF (20M) + HNO_3 (0.35M) + H_2SO_4 (1.1M)}.
 We immersed each \Ta tile in $170$ ml of this solution for $70$ s, then rinsed it with ultra-pure water for $30$ s and finally dried it with \ce{N_2} gas. This procedure removed between 4 and 5~$\mu$m of the surface (see Table\ \ref{tab:samples}).
 After the cleaning, the tiles were vacuum sealed into two nested plastic bags,
 of which the internal one remained also during the measurement (the absorption effect on $\gamma$ radiation is negligible).
 Since the cleaning operations could only be performed in the above-ground chemistry laboratory,
 we had to prevent that the cosmogenic activation of \ce{^{182}Ta} could spoil the effect of the ten-year-long storing deep-underground.
 We thus minimized the time spent above-ground by allowing only one tile at a time to leave the STELLA laboratory and cleaning each of the 6 tiles individually.
 In this way, we were able to ensure that no tile spent more than 30 minutes outside the Gran Sasso tunnel.
 
 After the chemical treatment described above, the total sample mass was $2015.12$ g (Table~\ref{tab:samples}), corresponding to $242.02$ mg of \ce{^{180m}Ta}.
 
 \section{Measurement}
 
 \begin{center}
  \begin{table}[t]
   \caption{Mass of the \Ta tiles before and after the cleaning treatment.
            The thickness reduction is estimated by assuming a uniform loss of material.
            The time spent outside the tunnel by each tile is also reported.
           }
   \begin{tabular}{l r r r r r}
    \hline  \\[-14pt] \hline
    Tile    &$m_\mathrm{in}$ [g]   &$m_\mathrm{fin}$ [g]   &$\Delta m$ [g]   &$\Delta x$ [$\upmu$m]   &$t_\mathrm{out}$ [min] \\
    \hline  \\[-8pt]
    1       &$339.75$              &$338.15$             &$1.60$           &$4.80$                  &$25$                   \\
    2       &$337.61$              &$336.22$             &$1.39$           &$4.16$                  &$20$                   \\
    3       &$341.72$              &$340.38$             &$1.34$           &$4.01$                  &$29$                   \\
    4       &$337.46$              &$335.93$             &$1.53$           &$4.58$                  &$29$                   \\
    5       &$333.05$              &$331.67$             &$1.38$           &$4.13$                  &$27$                   \\
    6       &$334.30$              &$332.77$             &$1.53$           &$4.58$                  &$27$                   \\
    \hline  \\[-14pt] \hline
   \end{tabular}
   \label{tab:samples}
  \end{table}
 \end{center}

 \begin{figure}[t]
  \centering
  \includegraphics[width=1.\columnwidth]{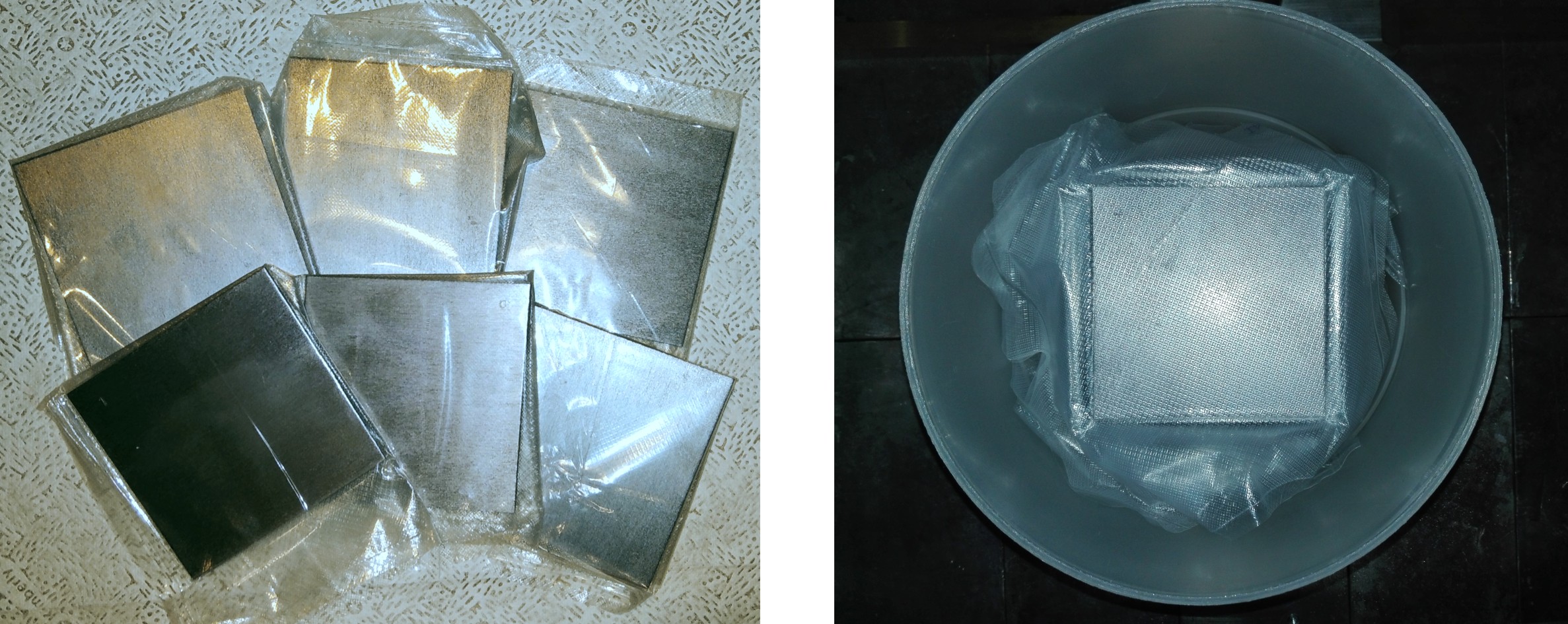}
  \caption{
           {\it (Left)} The \Ta sample consists of six tiles.
           {\it (Right)} The tiles arranged inside the Marinelli.
          }
  \label{fig:setup}
 \end{figure}
 
 The measurement of the \Ta sample was carried out with an \ULBHPGe detector in the STELLA laboratory~\cite{NEDER2000191,HEUSSER2006495}.
 The detector has a p-type germanium crystal with a volume of about $400$ cm$^3$; the counting efficiency has been optimized using a Marinelli-type geometry.
 The energy resolution measured with a high-statistics run is $(2.06 \pm 0.06)$ keV Full Width Half Maximum (FWHM) at the $1461$-keV peak of \ce{^{40}K}.
 
 To reduce the external background, the detector is encased in a multi-layer shield consisting of (from inside to outside) $5$ cm of electrolytic copper and low-radioactivity lead (30 cm from the bottom and 25 cm from the sides). The sample chamber has a volume of about 15 l ($250 \times 250 \times 240$ mm$^3$). The shield, together with the cryostat, is enclosed in an air-tight steel housing of 1-mm thickness, which is continuously flushed with highly-pure nitrogen gas in order to abate the radon-induced background.

 Figure~\ref{fig:setup} shows the \Ta-sample configuration during the measurement.
 The six tiles have been arranged inside a Marinelli-type beaker forming a cubic box around the detector end cap. Two tiles were placed on the top and a single tile on each lateral side.

 We acquired data over a $4$-yr period, from January 2019 to November 2022, for a total live-time of about $1.45$ yr.
 The measurement campaign comprises 13 runs of variable duration, from a few days up to three months (Table~\ref{tab:runs}).
 We could not operate continuously since the detectors of the STELLA laboratory are mainly devoted to radio-assay and screening of the materials to be employed in rare-event search experiments.
 For this reason we combined the collected runs into 4 datasets interspersed by the stop periods.
 All data have been acquired in the same sample-detector configuration and no significant difference between the datasets has been observed.

  \begin{table}[tb]
 \begin{center}
   \caption{Live time and start date of the runs analyzed in this work as divided into four major datasets.
            The discontinuities in the data acquisition are due to the fact that the STELLA facility is primarily assigned to the radio-assay and screening
            of materials.
            During the whole four-year period, the \Ta sample always remained inside the Marinelli-type beaker in the same configuration, constantly flushed with nitrogen.}
   \begin{tabular}{l r r}
    \hline  \\[-14pt] \hline
    Run             &Start    &\quad$t_\mathrm{live}$ [d] \\
    \hline  \\[-8pt]
    {\it Dataset I} \\
    \cline{1-1} \\[-8pt]
    \hphantom{1}1   &Jan 2019  &$ 43.0$ \\
    \hphantom{1}2   &Jun 2019  &$ 41.5$ \\
    \hphantom{1}3   &Aug 2019  &$ 59.3$ \\[+3pt]
    {\it Dataset II} \\
    \cline{1-1} \\[-8pt]
    \hphantom{1}4   &Feb 2020  &$ 58.9$ \\
    \hphantom{1}5   &Apr 2020  &$ 40.4$ \\[+3pt]
    {\it Dataset III} \\
    \cline{1-1} \\[-8pt]
    \hphantom{1}6   &May 2021  &$ 61.3$ \\
    \hphantom{1}7   &Jul 2021  &$  5.9$ \\
    \hphantom{1}8   &Jul 2021  &$100.6$ \\[+3pt]
    {\it Dataset IV} \\
    \cline{1-1} \\[-8pt]
    \hphantom{1}9   &Mar 2022  &$ 46.8$ \\
               10   &Jul 2022  &$ 29.7$ \\
               11   &Sep 2022  &$  5.9$ \\
               12   &Sep 2022  &$ 11.7$ \\
               13   &Oct 2022  &$ 22.5$ \\
    \hline  \\[-14pt] \hline
   \end{tabular}
   \label{tab:runs}
 \end{center}
  \end{table}
 
 \section{Data Analysis}
  
\begin{figure*}[tb]
   \centering
   \includegraphics[width=1.\textwidth]{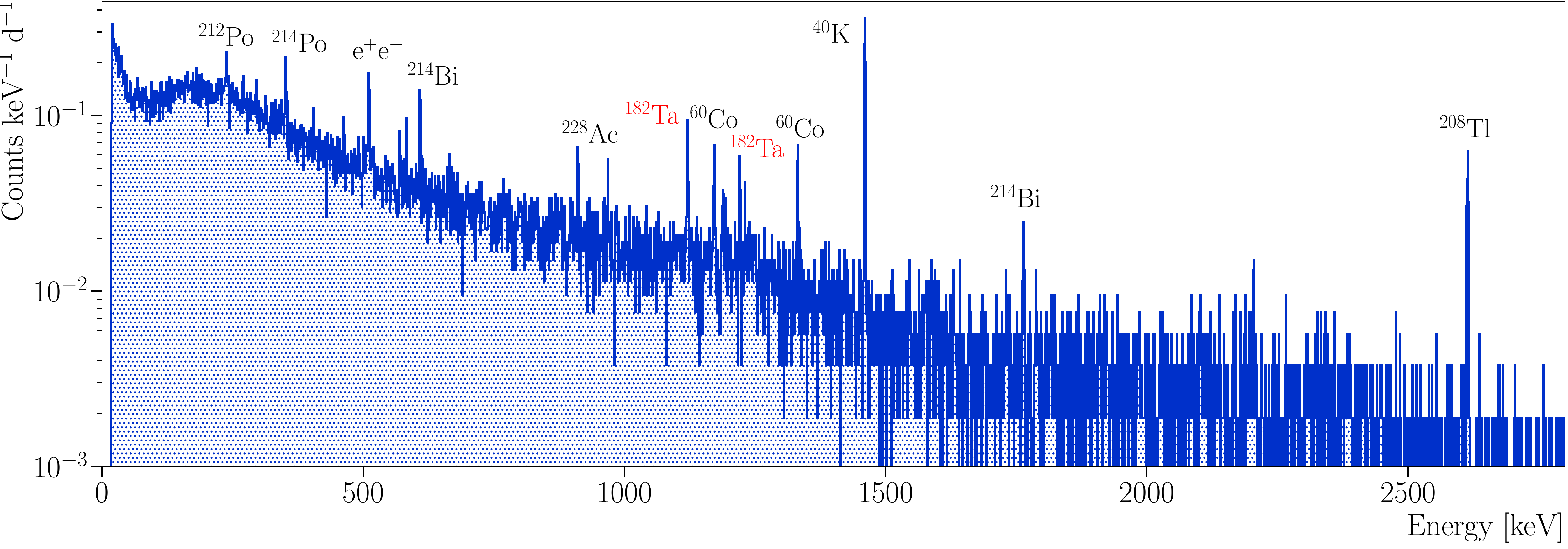}
   \caption{Sum spectrum of all the acquired runs merged after the calibration of the individual datasets.
            The total \Ta exposure is $2.911$ \kgyr, corresponding to $0.349$ \gyr of \mTa.
           }
   \label{fig:spectrum}
  \end{figure*}

 As a first step, we calibrated each dataset by referring to a set of background peaks from internal contamination of either the detector or the sample, namely
 \ce{^{214}Pb} ($351.9$~keV), \ce{^{60}Co} ($1332.5$~keV), \ce{^{40}K} ($1460.8$~keV) and \ce{^{208}Tl} ($2614.6$~keV);
 the detector response proved to be linear over the whole $(0-3)$-MeV range and stable during the measurement campaigns.
 The four individual calibrated spectra have then been rebinned and merged. The total spectrum is shown in Fig.~\ref{fig:spectrum}.
  
 From the analysis of the total spectrum, we were able to assess the internal contamination of the \Ta sample.
 Following Ref.~\cite{HEISEL2009741}, we calculated the activities of the various nuclides by reconstructing the detection efficiency of the full-energy peaks (FEPs)
 with MaGe, a Monte-Carlo simulation code based on the GEANT4 toolkit~\cite{Boswell:2011aaa}.
 The results are listed in Table~\ref{tab:contam}.
 As it can be seen, the sample is extremely radiopure and suitable for rare event searches, showing only \ce{^{40}K} and \ce{^{231}Pa} concentrations of some hundreds \uBqkg, the latter value being actually affected by a large uncertainty.
 In particular, the contribution from \ce{^{182}Ta} is subdominant to that of \ce{^{40}K} as a result of the ten-year-long storing deep-underground.
 The \ce{^{182}Ta} counting is not compatible with its radioactive decay; the measured $102$ \uBqkg is a weighted average over the whole four-dataset period. This residual activity might be due to a constant activation by the thermal-neutron component at STELLA. Given that the capture cross-section of \ce{^{181}Ta} is $20.5$ barn~\cite{Nakamura:2021Ta181}, this would correspond to a neutron flux of about $7.6 \times 10^{-6}$ \cmsqs, a value about an order of magnitude larger than the measured one~\cite{Best:2015yma}.\footnote{It is worth noting that the value reported in the reference has been measured in a different area of the underground LNGS facility.}
 
 The specific analysis of the \mTa decay is performed with two fits: the first corresponding to \betaM and \EC decay channels and the second for the $\gamma$ de-excitation / \IC (Fig.~\ref{fig:decay_scheme}).
 For each branch, we extract a single half-life parameter (\tGen) by combining a set of energy windows around the FEP of the $\gamma$-ray resulting from the metastable-state decay.
 In particular, for the \betaM mode, we consider one fit region around the FEPs at $234.0$ keV, while we do not include the one at $350.9$ keV since it almost fully overlaps with a prominent unconstrained FEP of \ce{^{214}Pb} at $351.9$ keV; we do not include the FEP at $103.6$ keV, given the much lower emission probability.
 This fit window also contains the low energy peak for the or the \EC branch, namely the FEP at $215.3$ keV. This ensures a common background parametrization for the two peaks of interest, namely $234.0$ keV and $215.3$ keV.
 Moreover, for the same fit we define a second window centred at the $332.3$ keV line of the \EC. We do not include the FEP at $93.3$ keV.
 Finally, for the $\gamma$ de-excitation / \IC mode, we consider a fit with only one region including both FEPs at $93.3$ keV and $103.6$ keV.
 The proper de-excitation FEPs of \mTa at $37.7$ keV and $39.5$ keV could not be investigated in this work since the combined detector's encapsulation and sample thickness result in detection efficiencies lower than $10^{-6}$.
 
 Each fit window $w$ contains multiple background $\gamma$ lines at energy $E_{b}$ and signal $\gamma$ lines at energy $E_{s}$.
 In general terms, the number of counts at energy $E$ are modeled as:
 \begin{multline}
  f_{w}(E) =  C_{w} + D_{w} E + \sum_{b} \frac{B_{w}}{\sqrt{2\pi}\sigma_{w}} e^{-(E-E_b)^{2} / 2\sigma^{2}_{w}} \\
  + \sum_{s} \frac{S_{w}}{\sqrt{2\pi}\sigma_{w}} e^{-(E-E_{s})^{2} / 2\sigma^{2}_{w}}
  \label{eq:model}
 \end{multline}
 where $C$ and $D$ are coefficients describing the polynomial component of the background, $B$ and $S$ are the number of counts in the background and signal peaks,
 while $\sigma$ is the energy resolution (assumed to be constant inside the window).
 The signal counts $S$ are connected to the decay half-life through the relation:
 \begin{equation}
  S (\tGen) = \text{ln\,2} \frac{\epsilon_{s}\ N_{A}\ \measT\ m\ \ia}{M} \frac{1}{\tGen},
 \end{equation}
 where $\epsilon_{s}$ is the detection efficiency of the specific signal $\gamma$ rays at $E_s$, $N_{A}$ is Avogadro's number, \measT is the measurement live-time ($\sim 1.45$~yr), $m$ is the total sample mass (2015.12~g),
 $M$ is the molar mass of \Ta (180.95) and \ia is the isotopic abundance of \mTa (0.000120).  
 
 \begin{center}
  \begin{table}[tb]
  \caption{Radionuclide concentrations in the \Ta sample measured with an \ULBHPGe detector. For each radionuclide, the activity value is obtained combining the results of the FEPs reported in the table. Measurements are quoted with the related combined expanded uncertainty, while limits are at 90\% C.\,L.\ .}
   \begin{tabular}{r r r r}
    \hline  \\[-14pt] \hline
   Chain          &\quad Nuclide      &Peaks                    &Activity       \\
                  &                   &\quad [keV]              &\quad [\uBqkg] \\[3pt]
   \hline  \\[-8pt]
   \ce{^{232}Th}  &\ce{^{228}Ra}      &$338.2, 911.2, 969.0$    & $ 42 \pm  10$ \\
                  &\ce{^{228}Th}      &$238.6, 583.0, 727.0$    & $ 55 \pm  10$ \\
                  &                   &$2614.6$                 &               \\[5pt]
   \ce{^{238}U}   &\ce{^{234}Th}      &$92.2,92.6$              & $< 140$       \\
                  &\ce{^{234m}Pa}     &$1001.0$                 & $< 650$       \\
                  &\ce{^{226}Ra}      &$241.9, 295.2, 351.9$    & $ 50 \pm   5$ \\
                  &                   &$609.3, 1120.3, 1764.5$  &               \\[5pt]
   \ce{^{235}U}   &\ce{^{235}U}       &$143.8, 185.7$           &$ <  10$       \\
                  &\ce{^{231}Pa}      &$300.1$,$302.7$          &$680 \pm 330$  \\
                  &\ce{^{227}Ac}      &$269.5$,$271.2$          &$88 \pm 33$    \\[5pt]
   \hline  \\[-8pt]
                  &\ce{^{40}K}        &$1460.8$                 & $570 \pm 180$ \\[5pt]
                  &\ce{^{137}Cs}      &$661.8$                  & $<   5$       \\[5pt]
                  &\ce{^{60}Co}       &$1173.2, 1332.5$         & $<   25$      \\[5pt]
                  &\ce{^{182}Ta}      &$1189.0, 1221.5, 1231.0$ &$102 \pm 15$\tableNoteI\\
    \hline  \\[-14pt] \hline
   \end{tabular}
   \begin{flushleft}
    \tableNoteI {\scriptsize This value is a weighted average of the four datasets.}
   \end{flushleft}
   \label{tab:contam}
  \end{table}
 \end{center}
  
 We perform a Maximum Likelihood fit using the Bayesian Analysis Toolkit (BAT) software package~\cite{Caldwell:2008fw}.
 The likelihood $\mathcal{L}$ is the product over each bin $i$ in each window $w$ of the Poisson probabilities that contribute to a specific decay branch:
 \begin{equation}
  \mathcal{L}(\textbf{p} \mid x) = \prod_{w}\prod_{i} \frac{f_{w,i}(\textbf{p})^{x_{i}}}{x_{i}!}e^{-f_{w,i}(\textbf{p})},
 \end{equation}
 where $\textbf{p}$ is the set of free parameters entering the fit and $x_{i}$ is the number of observed counts in the $i$-th bin.
 Referring to Eq.~\eqref{eq:model}, the free parameters in our case are:
 \begin{itemize}
  \item $C_{w}$ and $D_{w}$, set to flat prior probability;
  \item $B_{w}$, set to Gaussian prior if the nuclide activity can be inferred from Table~\ref{tab:contam}, or set to flat probability otherwise;
  \item $\sigma_{w}$, set to Gaussian prior. Its centroid and width are directly interpolated from the energy-resolution function extracted from dedicated calibration runs;%
   \footnote{
             We interpolated the energy-resolution function with the following polynomial function:
             $\sigma (E) = 0.504 + 3.4 \times 10^{-4} \ E - 5.7 \times 10^{-8} \ E^2$, where the energy $E$ is in keV.
            }
  \item $\tGen$, set to flat prior probability on the inverse, i.\,e.\ on the decay amplitude;
  \item $\epsilon_{s}$, set to Gaussian prior probability.
 \end{itemize}
 We constrain the fluctuation of the FEP positions by assuming a single energy-scale parameter for both signal and background peaks.
 For each decay mode, we assign a Gaussian prior to the peak position, centered on the nominal energy and taking its width as the uncertainty from the calibration functions. We assign the same energy-shift parameter to all peaks in the fit window.
 Finally, we estimate the efficiencies $\epsilon_{s}$ via Monte-Carlo simulations~\cite{Boswell:2011aaa} and assume a conservative systematic error of 7\%. 
 Uncertainties on the sample mass and isotopic abundance act as scaling parameters in the conversion of the signal strength into a limit on the half-life, however their effect on the final result is negligible.
 After marginalising the posterior probability distribution for the parameter of interest, i.\,e.\ the inverse of \tGen, we extract the 0.9 quantile representing the 90\% credibility interval (C.\,I.).
 This is the value we quote as a limit.

 \section{Results}
  
\begin{figure}[tb]
  \centering
  \includegraphics[width=1.\columnwidth]{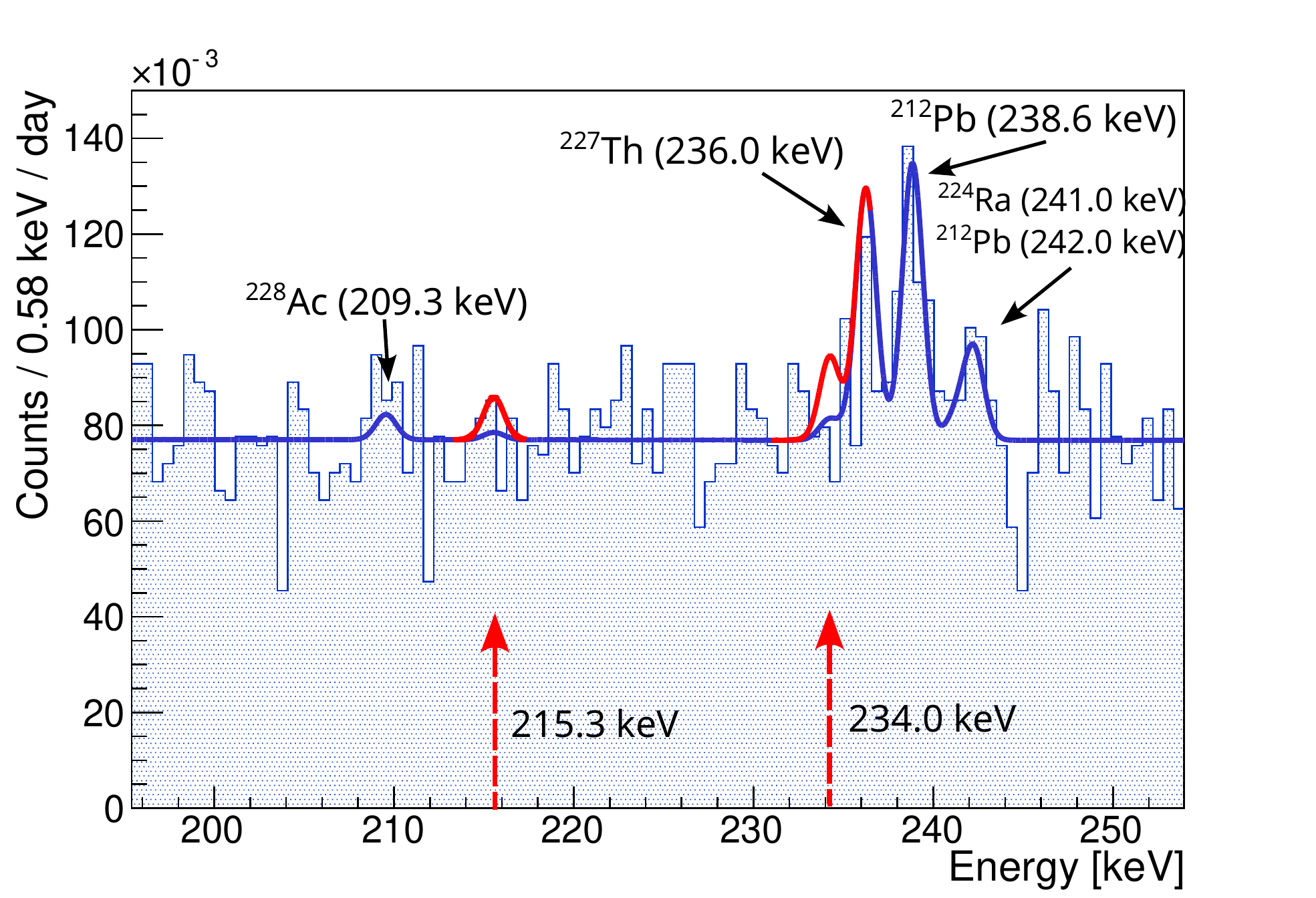}
  \includegraphics[width=1.\columnwidth]{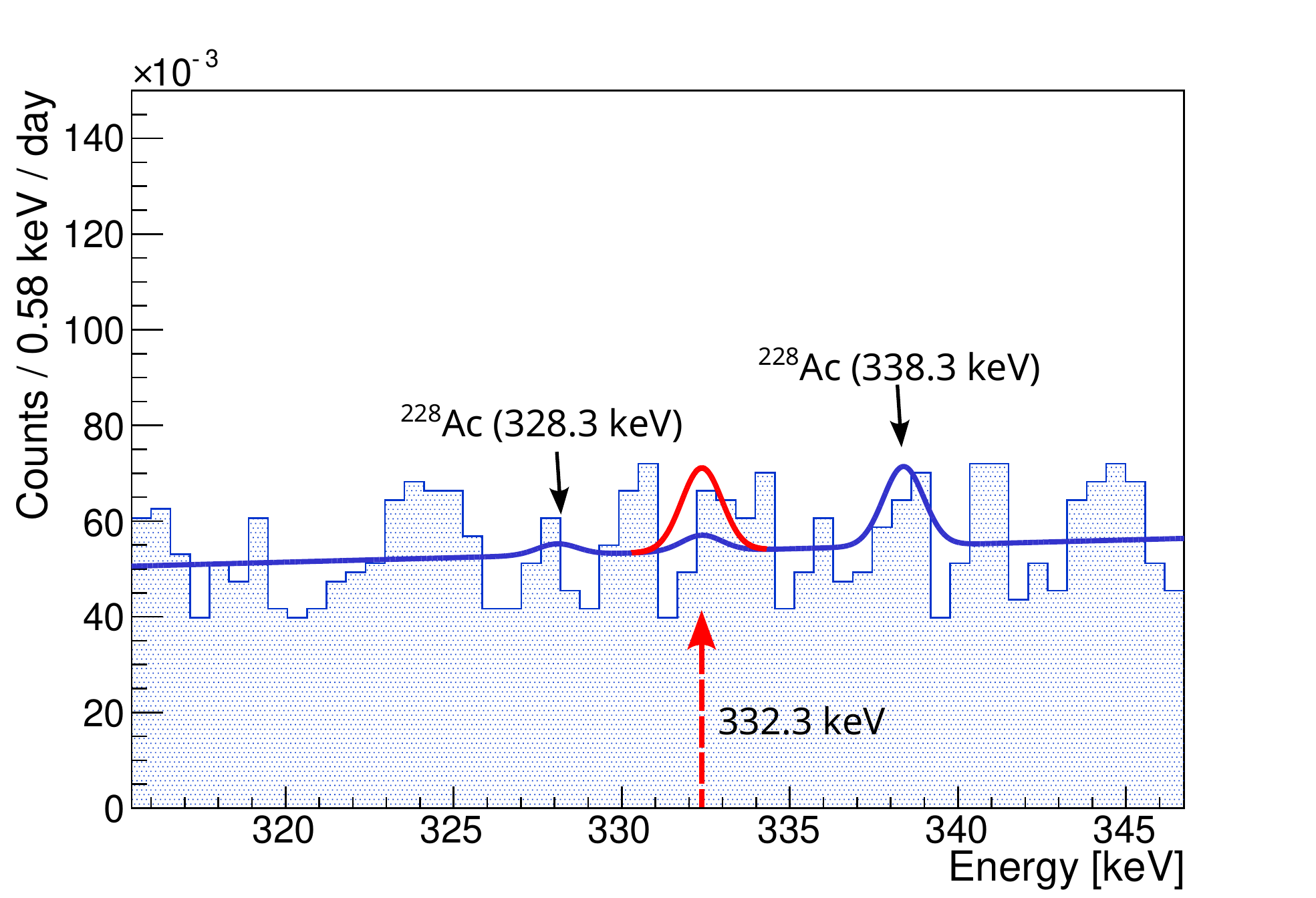}
  \caption{
           Fit windows for the \betaM and \EC decay branches of \mTa.
           The blue lines represent the best fit associated to the posterior mode;
           the red lines correspond to a signal peak with associated yield from the 90\% quantile of the inverse-half-life parameter posterior. The background and signal FEPs are indicated by the solid and dashed arrows, respectively.
          }
  \label{fig:BetaECFit}
 \end{figure}

\begin{figure}[tb]
  \centering
  \includegraphics[width=1.\columnwidth]{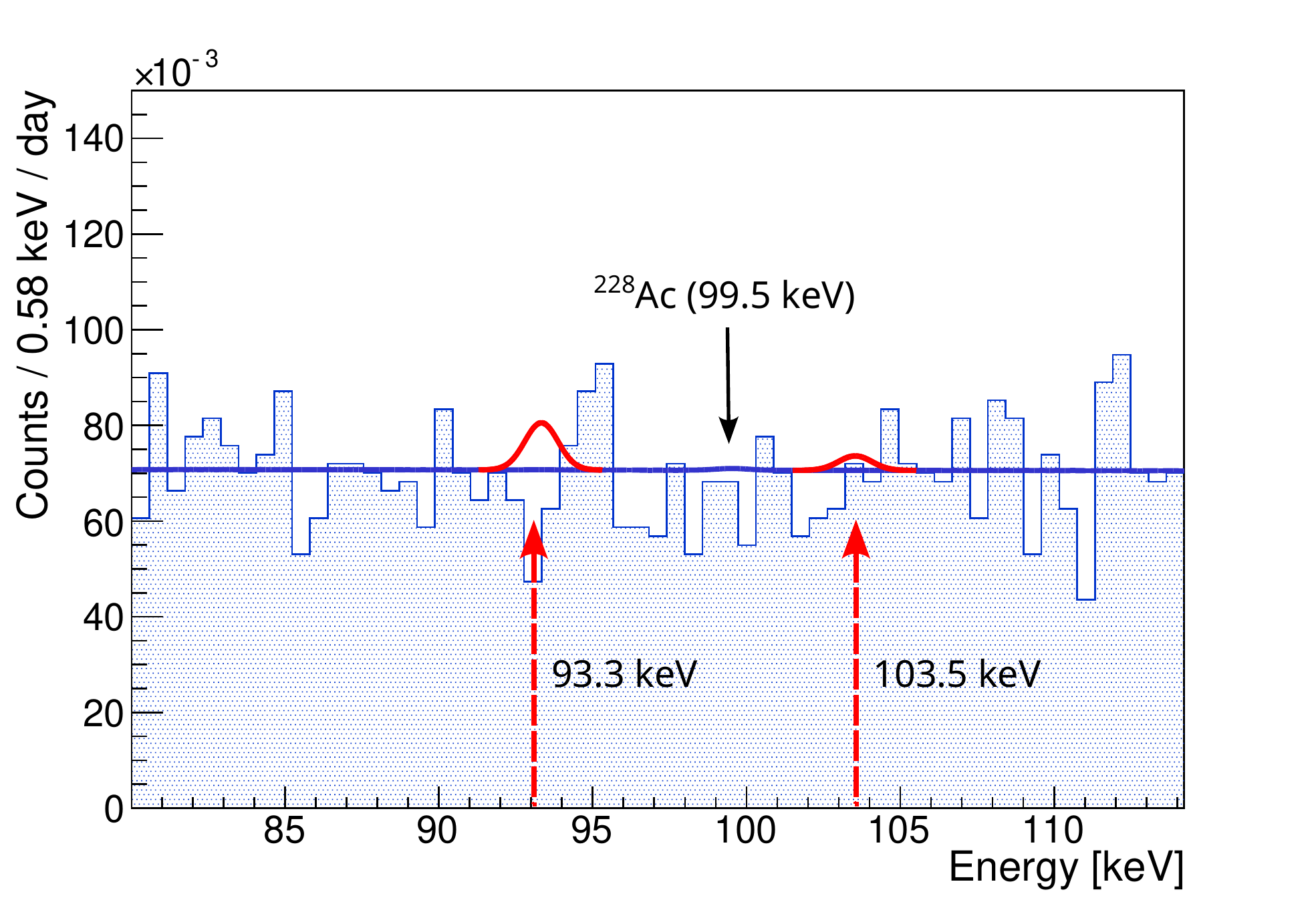}
  \caption{
           Fit window for the $\gamma$ de-excitation / \IC decay branch of \mTa.
           The blue line represents the best fit associated to the posterior mode;
           the red lines correspond to a signal peak with associated yield from the 90\% quantile of the inverse-half-life parameter posterior. The background and signal FEPs are indicated by the solid and dashed arrows, respectively.
          }
  \label{fig:DeExcFit}
 \end{figure}

The fits of the three decay branches are shown in Figs.~\ref{fig:BetaECFit} and \ref{fig:DeExcFit}, while the details of all the regions of interest (ROIs) are reported in Table~\ref{tab:ROIs}.
The \mTa peak counts are consistent with zero counts within $1\sigma$ in all cases and no signal has been observed; we therefore quote limits on the half-life of each decay channel.

 The first window of the \betaM and \EC fit (Figure~\ref{fig:BetaECFit}, top) also includes the background FEPs from  \ce{^{228}Ac} ($209.3$ keV),
 \ce{^{227}Th} ($236.0$ keV), \ce{^{212}Pb} ($238.6$ keV), \ce{^{214}Pb} ($242.0$ keV) and \ce{^{224}Ra} ($241.0$ keV).
 The second window (Figure~\ref{fig:BetaECFit}, bottom) contains background peaks from \ce{^{228}Ac} ($328.0$ keV and $338.3$ keV).
 The resulting limits are:
 \begin{align}
    \tb &> 1.1 \times 10^{18} \mbox{yr (90\% C.\,I.)} \\
    \tEC &> 1.6 \times 10^{18} \mbox{yr (90\% C.\,I.)}
 \end{align}
 which corresponds to an improvement by almost a factor $20$ for the \betaM branch and around one order of magnitude for the \EC decay with respect to the current most stringent limits~\cite{Lehnert:2016iku,Chan:2018chm}.
 For comparison purposes with other works we also can quote a combined limit by summing the two partial decay constants from which we extract: $\tECb > 6.5 \times 10^{17}$ yr $(90\%~\text{C.\,I.})$.
 
 Finally, Fig.~\ref{fig:DeExcFit} shows the fit of the $\gamma$ de-excitation~/~\IC mode.
 The same ROI includes the two FEPs at $93.3$ keV and $103.6$ keV, together with an expected background line from \ce{^{228}Ac} ($99.5$ keV).
 The lower half-life limit is
 \begin{equation}
    \tIC > 4.1 \times 10^{15} \mbox{yr (90\% C.\,I.)}, \label{eq:IC}
 \end{equation}
 gaining an improvement of a factor $30$ with respect to the current bound~\cite{Lehnert:2019tuw}.

  \begin{table*}[tb]
  \begin{center}
      
  \caption{
           Description of the fit windows. For each ROI, the detection efficiency $\epsilon_s$, the FWHM resolution, the background FEPs and the expected contribution $B$ are reported (refer to Eq.~\eqref{eq:model}).
          }
   \begin{tabular}{l r r r r r r}
    \hline  \\[-14pt] \hline
   Channel        &ROI         &\mTa FEP          &Efficiency                 &FWHM              &Bkg FEPs                      &Bkg counts    \\
                  &[keV]       &[keV]             &(\%)                       &[keV]             &[keV]                         &[\ckeVkgday]  \\[3pt]
   \hline  \\[-8pt]
   \betaM         &$195.3-254$   &$234.0$           &$1.56$                     &$1.37 \pm 0.01$   &$236.0, 238.6, 241.0, 242.0$  &$0.063$       \\[5pt]
   \EC            &$195.3-254$   &$215.3$           &$1.30$                     &$1.35 \pm 0.01$   &$209.3$                       &$0.063$       \\
                  &$318-347$   &$332.3$           &$2.70$                     &$1.44 \pm 0.01$   &$328.0, 338.3$                &$0.047$       \\[5pt]
   $\gamma$ / \IC & $80-115$   & $93.3$           &$0.0036$\tableNoteII       &$1.26 \pm 0.01$   &$99.5$                        &$0.06$        \\
                  &            &$103.5$           &$0.0011$\tableNoteII       &$1.26 \pm 0.01$   &$99.5$                        &$0.06$        \\[5pt]
    \hline  \\[-14pt] \hline
   \end{tabular}
   \begin{flushleft}
    \tableNoteII {\scriptsize This value also includes the branching ratio.}
   \end{flushleft}
   \label{tab:ROIs}
\end{center}
  \end{table*}

\section{Constraints on Dark Matter}

\begin{figure}[t]
   \centering
   \includegraphics[width=1.\columnwidth]{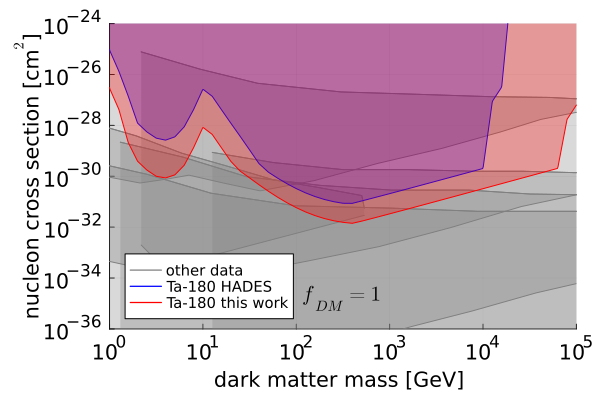}
   \includegraphics[width=1.\columnwidth]{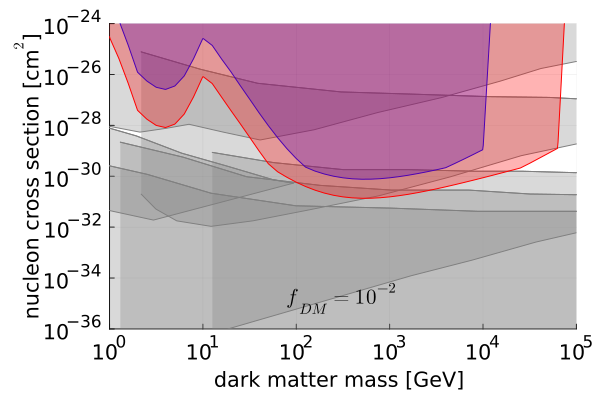}
   \includegraphics[width=1.\columnwidth]{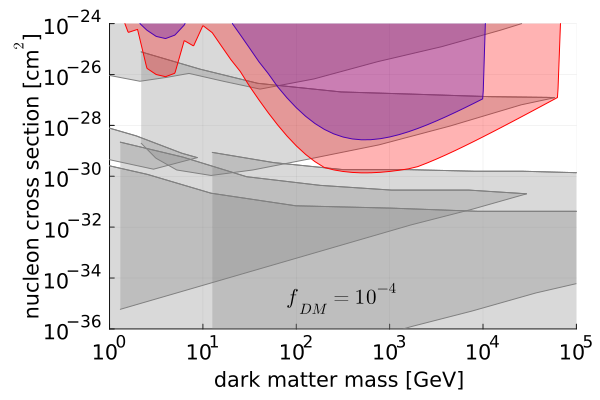}
   \caption{Exclusion plots for strongly-interacting DM. The three cases represent as many scenarios in terms of fraction of strongly-interacting            DM over the total DM density in the Solar System. From top to bottom: $100\%$, $1\%$ and $0.01\%$. The gray areas are the regions                previously excluded by other experimental searches (see Ref.~\cite{Lehnert:2019tuw} and references therein). The purple area indicates 
            the region currently excluded by the study of \mTa in HADES, while the red area that excluded in this work.
            The tighter limits on the $\gamma$ / \IC decay branch enhance the probing power to smaller nucleon-DM cross-sections.
            The greater rock overburden of LNGS extends the exclusion to higher masses.
            It can be seen that unexplored fractions of the parameter-space have now been probe for scenarios with a fraction of strongly-interacting DM of $1\%$ or less.
           }
   \label{fig:DMexclusion}
  \end{figure}

The study of \mTa is also valuable to constrain specific Dark Matter (DM) scenarios, as outlined in Ref.~\cite{Pospelov:2020} and demonstrated in Ref.~\cite{Lehnert:2019tuw}.
The general idea is that the nuclear energy stored in the metastable state of an isomer could be released due to DM interactions, so to create an experimental signature or to accelerate the DM particles. \mTa is of particular interest due its stability, allowing for experiments in low-background environments to be competitive. 

In strongly-interacting DM scenarios, the DM particles can interact with the overburden of an underground laboratory, hence slowing down from their initial galactic velocity distribution. Depending on the magnitude of the DM-nucleon cross-section and on the DM-particle mass, a full thermalization could take place in the overburden and in the end the DM particles might not have sufficient kinetic energy to produce detectable nuclear recoils in large-scale WIMP detectors. In this scenario, the thermalized DM could still interact with the metastable state of \mTa, leading to its de-excitation by absorbing the angular momentum of the $9^- \to 2^+$ transition. The experimental signature is the decay of the $^{180}$Ta ground state and it is identical to the $\gamma$ / \IC decay branches investigated in this work; therefore a positive observation of this decay would not be immediately indicative of a DM de-excitation. The smoking-gun for a DM-dominated de-excitation of \mTa would be an overburden-dependent half-life of the decay (scaling with the DM density) and would necessitate of multiple measurements in different underground laboratories. However, a non-observation can be used to constrain both the $\gamma$ / \IC decay branch and DM de-excitation simultaneously.  

Using the half-life limit for the $\gamma$ / \IC decay branch in Eq.~\eqref{eq:IC} and the same procedure outlined in Ref.~\cite{Lehnert:2019tuw}, we obtain the exclusion plots shown in Fig.~\ref{fig:DMexclusion}. In the figure, we represent three DM scenarios, corresponding to as many fractions of the strongly-interacting component over the total DM, namely $100\%$, $1\%$ and $0.01\%$. The limits coming from other experimental searches are rendered in gray, while the previous bounds from \mTa are shown in purple and the new limits from this work are shown in red. In the case where $100\%$ of the galactic DM density consists of strongly interacting DM (top panel), the parameter-space has already been probed by shallow-depth, surface, or air-borne DM detectors. Here, \mTa provides a uniquely different exclusion for this DM scenario. In the scenario where $1\%$ of the galactic DM density consists of strongly interacting DM (central panel), unconstrained regions exist; here we can exclude new parameter space, namely for masses above $10^{4}$ GeV. Finally, for a sub-component fraction of $0.01\%$, large areas of the parameter-space previously unconstrained, have now been excluded from $10^{2}$ to $10^{5}$ GeV. 

The main differences between the previous search with \mTa~\cite{Lehnert:2019tuw} and the current work are, on the one side, the improved half-life limit from $1.3 \times 10^{14}$~yr to $4.1 \times 10^{15}$~yr ($90\%$ C.\,I.); on the other, the increased overburden from $225$~m of HADES to $1400$~m of LNGS. The former factor enhances the probing power to smaller nucleon-DM cross-sections. The latter extends the exclusion to higher masses, as heavier DM particles are better thermalized in the overburden, thus increasing the DM density inside the detector.

We also considered other composite DM models based on \mTa measurements, as described in Ref.~\cite{Lehnert:2019tuw}. However, these have now been surpassed by measurements in other isotopes~\cite{Song2021}: the probed higher DM mass-splittings and the lower cross sections make \mTa not competitive anymore.

\section{Outlook and perspectives}
 
 The new limits on the decay of \mTa reported in this paper improve the previous ones by at least one order of magnitude. These results were obtained because of the excellent performance of the \ULBHPGe in the STELLA laboratory, combined with the reduction of the \ce{^{182}Ta} background to a subdominant level due to the ten-year-long storage deep-underground.

 There are different possible strategies to further improve the experimental sensitivity. The simplest way would be to increase the measurement time. In the current setup, a 5-year-long measurement would translate into a gain of a factor $\sim2$. However, in a background-limited detector, the sensitivity scales as $\tGen \propto \varepsilon \cdot B^{-1/2}$~\cite{Broerman:2020hfj}, therefore increasing the detection efficiency $\varepsilon$ by optimising the sample geometry would be more effective than reducing the background $B$.

 A valid  method to enhance the detection efficiency consists in using a HPGe detector array as adopted by the TGV collaboration~\cite{Rukhadze:2011zz}. There, thin foils of the sample material have been inserted between $16$ pairs of large-area HPGe detectors stacked in an array structure (in that case, they have been searching for different modes of double beta decay in $^{106}$Cd). By exploiting the coincidence between neighbouring face-to-face detectors, this method can lead to a strong background reduction (up to a factor $10$) along with an increase of the detection efficiency (by a factor $2)$. Alternatively, the same idea could be implemented by using rather thick \ce{Ta} samples sandwiched between pairs face-to-face \ce{Ge} detectors stacked in multiple towers, as it is being currently done with the {\sc Majorana} Demonstrator, where $17.4$ kg of \Ta are allocated in between $23$ HPGe detectors~\cite{Majorana:ta180m}.

 A further modification of the stacked detector-array approach would be to use a tower of \ce{Ge} wafers working as cryogenic calorimeters, similar to the light detectors widely utilized in the CUPID-0~\cite{CUPID:2018kff} or CUPID-Mo~\cite{Armengaud:2019loe} experiments. In this approach, \ce{Ta} metal foils and Ge wafers (both with optimized thickness) could be stacked minimizing the distance between detectors and sources. The typical energy resolution for cryogenic light detectors is about $100$ eV FWHM~\cite{Artusa:2016maw}, and could be further pushed down to about $20$ eV by exploiting the Neganov-Trofimov-Luke amplification~\cite{Berge:2017nys}. Such an improved energy resolution would help to minimize background contributions to the region of interest, leading to a factor $\sim 5$ enhancement in the experimental sensitivity.

 Half-lives larger than $10^{21}$ yr could be in principle probed by performing a calorimetric measurement -- i.\,e.\ with the source embedded in the detector -- by including \Ta into a compound suitable for the use as a scintillator or as a cryogenic calorimeter. Promising materials are for example \ce{LiTaO_3}, \ce{Cs_2TaCl_6}~\cite{tehrani2023charge} or metallic tantalum itself, being a superconductor~\cite{Hochberg:2015fth, Billard:2016giu, Pattavina:2019pxw, Ricochet:2021rjo}. This approach would especially benefit searches for low-energy $\gamma$'s, where the improvement in the detection efficiency would be multiple orders of magnitude.

 An entirely different approach are indirect searches for increased concentrations of the daughter isotopes $^{180}$W and $^{180}$Hf in geological tantalum samples \cite{Lehnert_2021}. Tantalite and columbite minerals can be more than $10^9$~yr old, allowing for ample time of daughter accumulation. The sensitivity of these geological searches is difficult to predict, crucially depending on trace impurities of tungsten and hafnium in the minerals as well as on the precision and sensitivity of mass spectrometry. 

 Despite the non-observation of \mTa decay thus far, a number of techniques now exist to perform these investigations. With adaptation and tuning of the methods described above, the theoretical predictions are within reach. An observation of one or even multiple decay modes of \mTa would not only be intriguing for understanding the nuclear physics of the longest lived nuclear isomer but also opens an avenue for studying certain dark matter scenarios if the decay is observed in multiple underground setups with different overburden.

\section*{Acknowledgements} 
 
 We thank the director and staff of the Laboratori Na-zionali del Gran Sasso for the support.
 S.\,S.\,N.\ is supported by the Arthur B.\ McDonald Canadian Astroparticle Physics Research Institute.
 L.\,P.\ research activities are supported by European Union's Horizon 2020 research and innovation program under the Marie Sk\l o-dowska-Curie grant agreement N.\ 101029688.
 We thank and acknowledge Harikrishnan Ramani for fruitful discussions and his excellent help in calculating the dark matter exclusions.
 
\section*{Availability of data and material}

Data will be made available on reasonable request to the corresponding author.

 \bibliography{ref}

\end{document}